\documentclass[preprint]{elsart}

\usepackage{graphicx}
\usepackage{amssymb}
\usepackage{bm}

\begin{document}

\begin{frontmatter}




\title{Impurity scattering effect on the zero-energy peak of the local density of states in a multi-quantum vortex core}


\author[A,B]{Takashi Yamane\corauthref{cor1}}
\ead{soraumi$\mbox{\_}$riku@pe.osakafu-u.ac.jp}
\author[C]{Yuki Nagai}
\author[D]{K. Tanaka}
\author[B]{Nobuhiko Hayashi}

\address[A]{Department of Physics and Electronics, Osaka Prefecture University, 
1-1 Gakuen-cho, Naka-ku, Sakai 599-8531, Japan}
\address[B]{NanoSquare Research Center (N2RC), Osaka Prefecture University, 1-2 Gakuen-cho, Naka-ku, Sakai 599-8570, Japan}
\address[C]{CCSE, Japan Atomic Energy Agency, 
5-1-5 Kashiwanoha, Kashiwa, Chiba 277-8587, Japan}
\address[D]{Department of Physics and Engineering Physics, University of Saskatchewan, 
116 Science Place, Saskatoon, Saskatchewan, S7N 5E2 Canada}

\corauth[cor1]{Corresponding author. Tel \& Fax: +81-72-254-8203. \\
N2RC, Osaka Prefecture University, C10 Bldg., 1-2 Gakuen-cho, Naka-ku, Sakai 599-8570, Japan.
}

\end{frontmatter}
\newpage
\begin{frontmatter}

\begin{abstract}
We theoretically study a non-magnetic impurity effect on the vortex bound states of a multi-quantum vortex.
The zero-energy peak of the local density of states is investigated for vortex cores with the winding numbers 2 and 4 within the framework of the quasiclassical theory of superconductivity.
We find that the zero-energy peaks, which appear away from the vortex center in the clean limit, move towards the vortex center with increasing the impurity scattering rate, resolving a contradiction between an experimental result and previous theoretical predictions.
\end{abstract}


\end{frontmatter}


\section{Introduction}
Much attention has been focused on the vortex core structure in superconductors.
A vortex with a multiple winding number is called the multi-quantum vortex or the giant vortex,
which accommodates multiple flux quanta.
While only vortices with the winding number $L=1$ are formed in usual situations,
a multi-quantum vortex with $L>1$ may be realized in a small-size superconductor because of a geometrical restriction of the split into individual vortices with $L=1$ \cite{Schweigert98}.
Recently, an observation of multi-quantum vortices by scanning tunneling microscopy (STM)
was reported, where the experiment was performed on nanometer-sized Pb islands \cite{Cren11}. 
There is, however, a contradiction between the experimental observation \cite{Cren11} and previous theoretical predictions \cite{Tanaka93,Volovik93,Rainer96,Virtanen99,Virtanen00,Melnikov02,Tanaka02}.
In the STM experiment, the zero-bias peak of the tunneling conductance was observed
at the vortex center for both odd and even winding-number vortices \cite{Cren11}.
In contrast, previous theories predicted that the zero-energy peak of the local density of states exists at the vortex center only for odd winding-number vortices and not for even ones \cite{Tanaka93,Volovik93,Rainer96,Virtanen99,Virtanen00,Melnikov02,Tanaka02}.
Therefore, it is necessary to resolve this contradiction for even winding-number cases.

In this paper, we claim that an impurity scattering effect can reconcile the above contradiction.
The zero-energy local density of states inside multi-quantum vortex cores with $L=2$ and $4$
is investigated under the influence of impurities.
The zero-energy peaks appear away from the vortex center in the clean limit,
which is consistent with previous theories \cite{Tanaka93,Volovik93,Rainer96,Virtanen99,Virtanen00,Melnikov02,Tanaka02}.
We find that these peaks move towards the vortex center with increasing the impurity scattering rate,
resulting in a peak situated at the vortex center.
The experimental result \cite{Cren11}, which seemingly contradicts previous theories,
can be explained by this impurity scattering effect.

\section{Formulation}
We consider a single vortex in a single-band $s$-wave superconductor.
The system is assumed to be an isotropic two-dimensional layer perpendicular to the vorticity along the $z$ axis.
In a circular coordinate system within the layer,
the real-space position is ${\bf r}=(r\cos\phi,r\sin\phi)$, and
the unit vector ${\bar{\bf k}}$
represents the sense of the wave number
on a Fermi surface assumed to be circular.
The Fermi velocity is ${\bf v}_{\rm F}=v_{\rm F}{\bar{\bf k}}$.

The pair potential $\Delta({\bf r})$ around a multi-quantum vortex with winding number $L$ is
represented as
\begin{eqnarray}
\Delta({\bf r})
=
{\bar \Delta}(r) e^{i L\phi},
\label{eq:pair}
\end{eqnarray}
where ${\bar \Delta}(r)$ is the pair potential amplitude
and
the vortex center is situated at ${\bf r}=0$.
The vortex core structure is self-consistently calculated by means of the quasiclassical theory of superconductivity \cite{QC}.
The Eilenberger equation is numerically solved to obtain the quasiclassical Green's function ${\hat g}(i\omega_n,{\bf r},{\bar{\bf k}})$.
The effect of impurities distributed randomly in the system is taken into account through the impurity self energy ${\hat \Sigma}(i\omega_n,{\bf r},{\bar{\bf k}})$.
These quantities and
the Eilenberger equation to be solved are \cite{QC,Hayashi05,Hayashi12}
\begin{eqnarray}
{\hat g}=
-i\pi
\pmatrix{
g &
if \cr
-if^{\dagger} &
-g \cr
}, \quad
{\hat \Sigma}
&=&
\pmatrix{
\Sigma_{\rm d} &
\Sigma_{12} \cr
\Sigma_{21} &
-\Sigma_{\rm d} \cr
},
\end{eqnarray}
\begin{eqnarray}
i {\bm v}_{\rm F} \cdot
{\bm \nabla}{\hat g}
+ \bigl[ i{\tilde \omega}_n {\hat \tau}_{3}-\hat{\tilde \Delta},
{\hat g} \bigr]
&=& 0.
\label{eq:eilen2}
\end{eqnarray}
The equation is supplemented by the normalization condition
${\hat g}^2=-\pi^2 {\hat \tau}_0$.
Here, ${\hat \tau}_3$ is the $z$ component of  the Pauli matrix,
${\hat \tau}_0$ is the unit matrix,
and
the brackets denote the commutator $[{\hat A},{\hat B}]={\hat A}{\hat B}-{\hat B}{\hat A}$.
The Eilenberger equation contains  the renormalized Matsubara frequency ${\tilde \omega}_n$
and pair potential $\hat{\tilde \Delta}$ \cite{Hayashi05},
\begin{equation}
i{\tilde \omega}_n = i\omega_n - \Sigma_{\rm d}, \qquad
\hat{\tilde \Delta}=
\pmatrix{
0 &
\Delta + \Sigma_{12} \cr
- (\Delta^{*} - \Sigma_{21}) &
0 \cr
}.
\end{equation}
We have neglected the vector potential assuming the case of a large Ginzburg-Landau parameter.
Throughout the paper, we use units in which $\hbar = k_{\rm B} = 1$.

Considering an $s$-wave non-magnetic impurity scattering and the $t$-matrix,
the impurity self energy ${\hat \Sigma}$ is given by \cite{Hayashi12}
\begin{eqnarray}
{\hat \Sigma}(i\omega_n, {\bf r})
=
\frac{\Gamma_{\rm n} }{1-(\sin^2\delta_0)(1-C)}
\pmatrix{
-i \langle g \rangle &
   \langle f \rangle \cr
 - \langle f^{\dagger} \rangle &
 i \langle g \rangle \cr
},
\label{eq:imp-self2}
\end{eqnarray}
where $C=\langle g \rangle^2 +  \langle f \rangle \langle f^\dagger \rangle$
with $\langle \cdots \rangle$ being
the average over the Fermi surface with respect to ${\bar{\bf k}}$.
The impurity scattering rate in the normal state is
$\Gamma_{\rm n}$, which is related to the mean free path
$l=v_{\rm F}/2\Gamma_{\rm n}$.
The scattering phase shift is $\delta_0$.
In the present study,
we set $\delta_0=0$ keeping $\Gamma_{\rm n}$ finite,
which corresponds to the Born limit.

   The self-consistency equation for $\Delta$,
called the gap equation, is given as
\begin{equation}
\Delta({\bf r})
=\lambda \pi T
\sum_{-\omega_{\rm c} < \omega_n < \omega_{\rm c}}
\Bigl\langle f(i\omega_n, {\bf r},{\bar {\bf k}})
\Bigr\rangle,
\label{eq:gap}
\end{equation}
where
$\omega_{\rm c}$ is the cutoff energy
and the coupling constant $\lambda$ is given by
\begin{equation}
\frac{1}{\lambda}
=
\ln\Bigl(\frac{T}{T_{{\rm c}}} \Bigr)
+ \sum_{0 \le n < (\omega_{\rm c}/\pi T -1)/2}   \frac{2}{2n+1}.
\label{eq:coupling}
\end{equation}
 Here, 
$T$ is the temperature
and
$T_{{\rm c}}$ is the superconducting critical temperature.
We set $\omega_{\rm c}=10\Delta_0$
with $\Delta_0$ being the BCS pair-potential amplitude at zero temperature.

The Eilenberger equation, the impurity self energy, and the gap equation
are numerically solved self-consistently at $T=0.1 T_{{\rm c}}$
to determine the spatial profile of the pair potential amplitude ${\bar \Delta}(r)$
following a procedure described in Ref.~\cite{Hayashi12}.
Then, using the pair potential determined self-consistently, the Eilenberger equation
and the impurity self energy are self-consistently solved
for a real energy $E$ with replacing $i\omega_n \to E+i\eta$ \cite{Kato02,Tanuma09}.
Here, $\eta$ is a small positive energy, which is technically necessary for obtaining numerically
a retarded Green's function.
We set $\eta=0.03 \Delta_0$.
The local density of states $N({\bf r}, E)$ is calculated as
\begin{equation}
N({\bf r},E)
=N_{\rm F}
\mbox{Re}
\Bigl\langle g(i\omega_n \to E+i\eta, {\bf r},{\bar {\bf k}})
\Bigr\rangle,
\label{eq:LDOS}
\end{equation}
where $N_{\rm F}$ is the density of states at the Fermi level in the normal state.

In the next section, we will show results for the pair potential and the zero-energy local density of states around the multi-quantum vortices with $L=2$ and $4$.
We define the coherence length $\xi_0=v_{\rm F}/\Delta_0$
and use it as the unit of the length.


\section{Result and Discussion}
In Fig.~\ref{fig:1}, we show the pair potential amplitudes ${\bar \Delta}(r)$
around the multi-quantum vortices with the winding numbers $L=2$ and $4$.
In the clean limit ($\Gamma_{\rm n}=0$),
the spatial profiles are consistent with results shown in Ref.~\cite{Virtanen99}.
A healing length, which is a length necessary for the pair potential to recover to a balk value,
decreases with increasing $\Gamma_{\rm n}$.
In other words, the vortex core radius shrinks with increasing $\Gamma_{\rm n}$.

In Fig.~\ref{fig:2}, shown are
numerical results for the zero-energy local density of states $N(r, E=0)$
around the multi-quantum vortices with $L=2$ and $4$.
They are plotted as a function of the radial distance $r$ from the vortex center
for several values of the impurity scattering rate $\Gamma_{\rm n}$.
In the clean limit ($\Gamma_{\rm n}=0$),
the zero-energy peaks appear at finite distances of the order of the coherence length,
which is consistent with previous theories \cite{Tanaka93,Volovik93,Rainer96,Virtanen99,Virtanen00,Melnikov02,Tanaka02}.
We find that the zero-energy peaks gradually move towards the vortex center
with increasing $\Gamma_{\rm n}$.
The two peaks reduce to a single one in the case of $L=4$.
In a dirty case ($\Gamma_{\rm n}=6 \Delta_0$),
the zero-energy peak is located almost at the vortex center $r=0$
and the peak height is equal to the normal-state density of states at the Fermi level $N_{\rm F}$.

We can make a physical interpretation of the result in the dirty state as follows.
When the quasiparticles inside a vortex core are frequently scattered by impurities,
they lose the information on the phase of the pair potential that they feel. 
As a consequence, the difference of the winding number $L$ related to the pair potential phase
becomes ineffective for the electronic structure of those quasiparticles
and 
the so-called normal core, as in $L=1$, is realized in the dirty state irrespective of value of $L$.
(Note that the pair potential itself keeps its winding number even in the dirty state,
namely  the winding number is kept being $L \neq 0$ and the vortex never disappears.)
At the normal core, the value of the local density of states 
is the normal-state density of states $N_{\rm F}$.
On the other hand, far away from a vortex, namely in the bulk,
the superconducting gap is fully opened in an $s$-wave superconductor
even in the dirty state
and the zero-energy density of states is zero.
As a result, the zero-energy local density of states has the value of $N_{\rm F}$
at the vortex core and decays to zero far away from the core
irrespective of winding number $L$.
Such a spatial variation of the zero-energy local density of states
exhibits an appearance of a peak structure as seen in Fig.~\ref{fig:1} 
(a dirty case of $\Gamma_{\rm n}=6 \Delta_0$).

In the STM experiment,
the superconducting Pb islands on a Si substrate are in the dirty state \cite{Cren11,Nishio08}.
Therefore, the absence rule of the zero-energy peak at the vortex center
for multi-quantum vortex with even winding number predicted theoretically in the clean limit \cite{Tanaka93,Volovik93,Rainer96,Virtanen99,Virtanen00,Melnikov02,Tanaka02}
is not applicable in that experiment \cite{Cren11}.
The presence of the zero-energy peak at the vortex center for $L=2$
observed experimentally \cite{Cren11} can naturally be understood in terms of the impurity scattering effect
discussed above.

\section{Conclusion}
We investigated the zero-energy density of states around the multi-quantum vortices
with the winding numbers $L=2$ and $4$ in the $s$-wave superconductor
by means of self-consistent numerical calculations based on the quasiclassical theory of superconductivity.
We found that the zero-energy peaks, which appear away from the vortex center in the clean limit, gradually move towards the vortex center with increasing the impurity scattering rate.
This result of the impurity scattering effect can explain
the presence of the zero-energy peak at the vortex center for $L=2$
observed experimentally \cite{Cren11}.
Effects of a finite system size \cite{Nagai12} and of inclusion of the vector potential
are left for future studies.

\section*{Acknowledgments}
We thank
N. Nakai,
Y. Hasegawa,
D. Roditchev,
H. Suematsu,
and
Y. Higashi
for helpful discussions.






\begin{thebibliography}{00}

\bibitem{Schweigert98}
V. A. Schweigert, F. M. Peeters, P. Singha Deo,
Phys. Rev. Lett. 81 (1998) 2783.

\bibitem{Cren11}
T. Cren, L. Serrier-Garcia, F. Debontridder, D. Roditchev,
Phys. Rev. Lett. 107 (2011) 097202.

\bibitem{Tanaka93}
Y. Tanaka, A. Hasegawa, H. Takayanagi,
Solid State Commun. 85 (1993) 321.

\bibitem{Volovik93}
G. E. Volovik,
JETP Lett. 58 (1993) 455.

\bibitem{Rainer96}
D. Rainer, J. A. Sauls, D. Waxman,
Phys. Rev. B 54 (1996) 10094.

\bibitem{Virtanen99}
S. M. M. Virtanen, M. M. Salomaa,
Phys. Rev. B 60 (1999) 14581.

\bibitem{Virtanen00}
S. M. M. Virtanen, M. M. Salomaa,
Physica B 284-288 (2000) 741.

\bibitem{Melnikov02}
A. S. Mel'nikov, V. M. Vinokur,
Nature 415 (2002) 60.

\bibitem{Tanaka02}
K. Tanaka, I. Robel, and B. Jank\`o,
Proc. Natl. Acad. Sci. USA 99 (2002) 5233.


\bibitem{QC}
G. Eilenberger, Z. Phys. 214 (1968) 195;
A. I. Larkin, Yu. N. Ovchinnikov, Sov. Phys. JETP 28 (1969) 1200;
J. W. Serene, D. Rainer, Phys. Rep. 101 (1983) 221.

\bibitem{Hayashi05}
N. Hayashi, Y. Kato, M. Sigrist, 
J. Low Temp. Phys. 139 (2005) 79.

\bibitem{Hayashi12}
N. Hayashi, Y. Higashi, N. Nakai, H. Suematsu,
to be published in Physica C [DOI: 10.1016/j.physc.2012.03.038].

\bibitem{Kato02}
Y. Kato, N. Hayashi,
J. Phys. Soc. Jpn. 71 (2002) 1721.

\bibitem{Tanuma09}
Y. Tanuma, N. Hayashi, Y. Tanaka, A. A. Golubov,
Phys. Rev. Lett. 102 (2009) 117003.

\bibitem{Nishio08}
T. Nishio, T. An, A. Nomura, K. Miyachi, T. Eguchi, H. Sakata, S. Lin,
N. Hayashi, N. Nakai, M. Machida, Y. Hasegawa,
Phys. Rev. Lett. 101 (2008) 167001.


\bibitem{Nagai12}
Y. Nagai, K. Tanaka, N. Hayashi,
Phys. Rev. B 86 (2012) 094526.

\end{thebibliography}




\newpage
\begin{figure}
\caption{  \label{fig:1}
Spatial profiles of the pair potential amplitude ${\bar \Delta}(r)$
around the multi-quantum vortices with the winding numbers (a) $L=2$ and (b) $L=4$
for several values of the impurity scattering rate $\Gamma_{\rm n}$.
The horizontal axis $r$ is the distance from the vortex center.
}
\end{figure}

\begin{figure}
\caption{  \label{fig:2}
Spatial profiles of the zero-energy local density of states $N(r, E=0)$
around the multi-quantum vortices with the winding numbers (a) $L=2$ and (b) $L=4$
for several values of the impurity scattering rate $\Gamma_{\rm n}$.
The horizontal axis $r$ is the radial distance from the vortex center.
}
\end{figure}

\begin{figure}
\vspace{10cm}
\end{figure}


\newpage
\begin{figure}[!t]
Figure 1(a) \\
 \\
 \\
 \\
 \\
 \\
 \\
\includegraphics[scale=1]{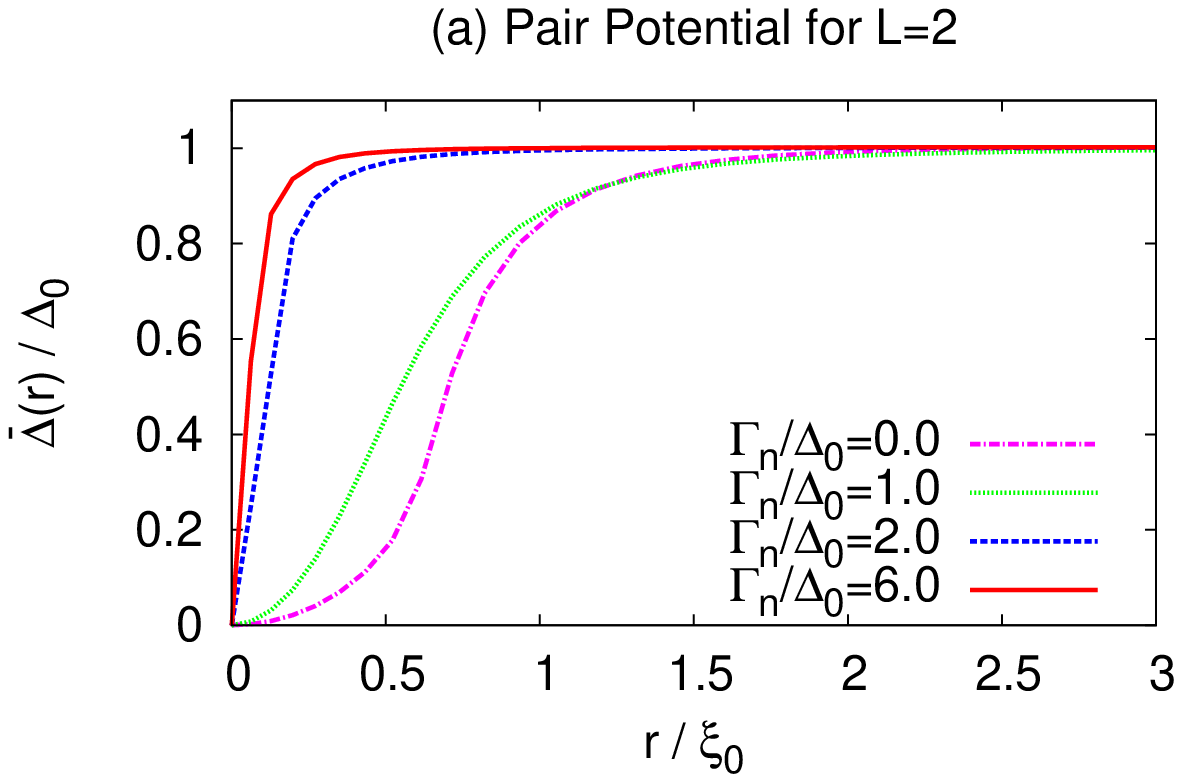}
\end{figure}

\newpage
\begin{figure}[!t]
Figure 1(b) \\
 \\
 \\
 \\
 \\
 \\
 \\
\includegraphics[scale=1]{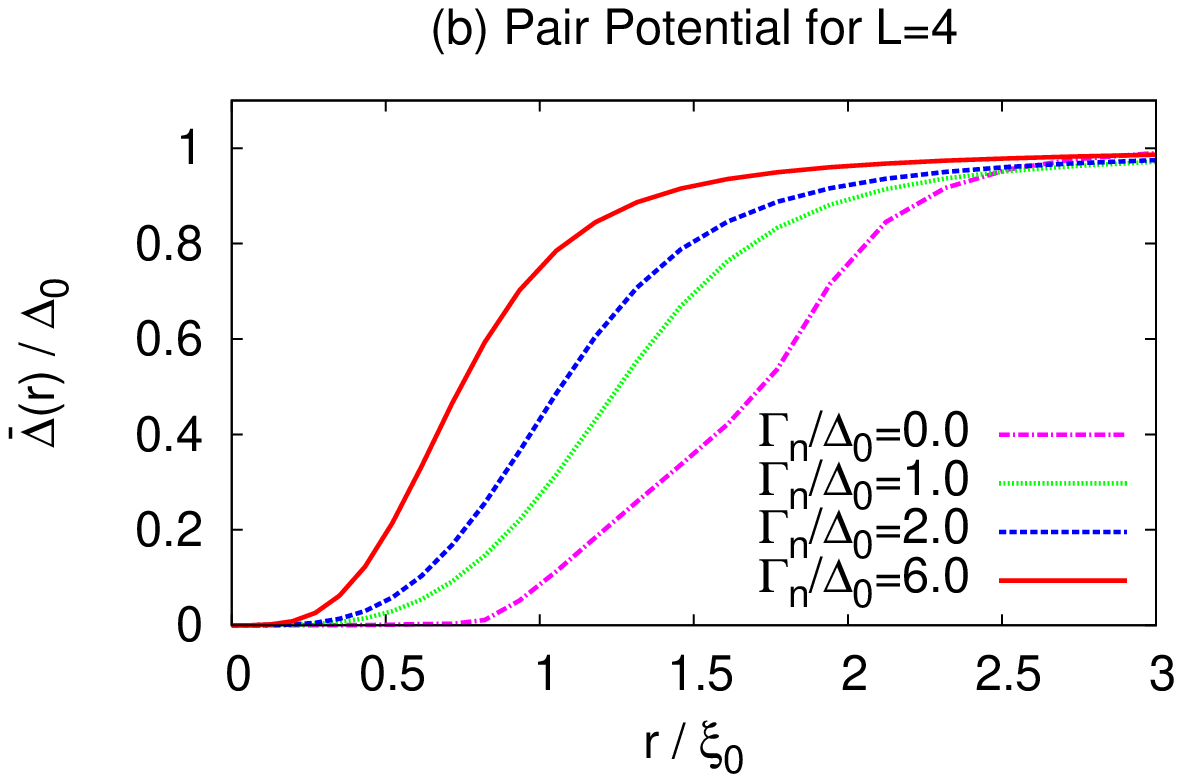}
\end{figure}

\newpage
\begin{figure}[!t]
Figure 2(a) \\
 \\
 \\
 \\
 \\
 \\
 \\
\includegraphics[scale=1]{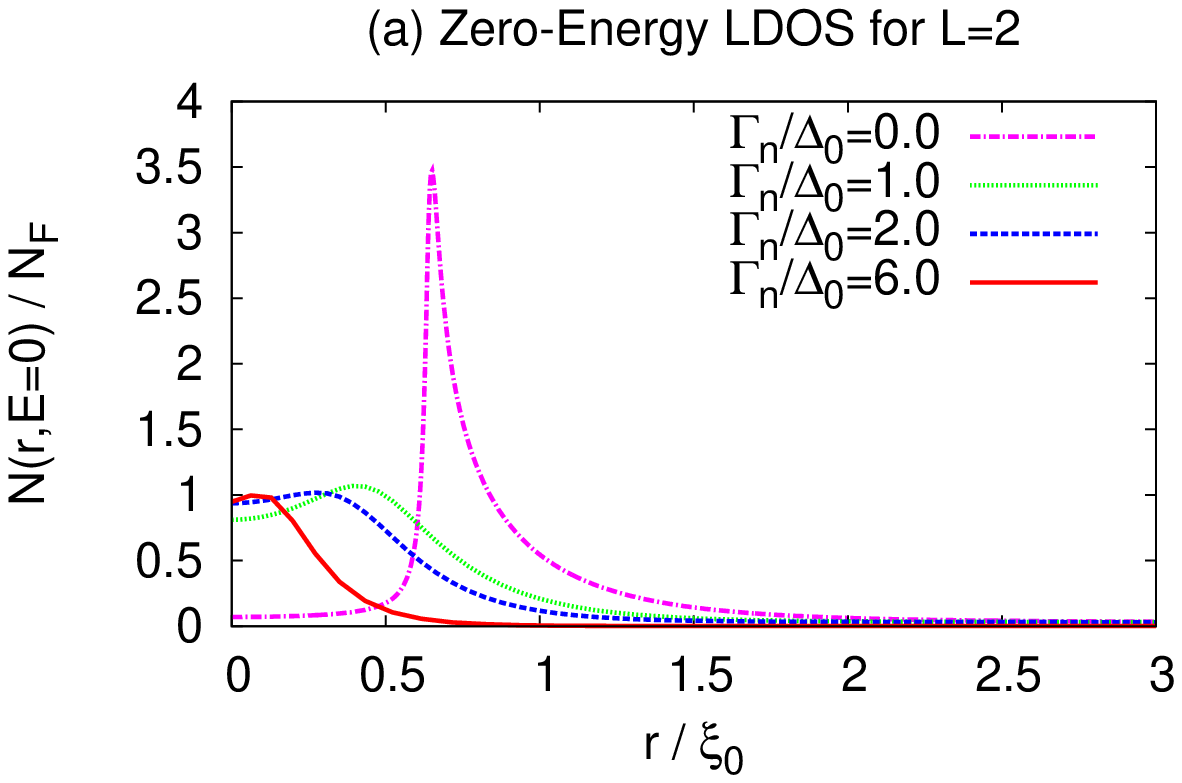}
\end{figure}

\newpage
\begin{figure}[!t]
Figure 2(b) \\
 \\
 \\
 \\
 \\
 \\
 \\
\includegraphics[scale=1]{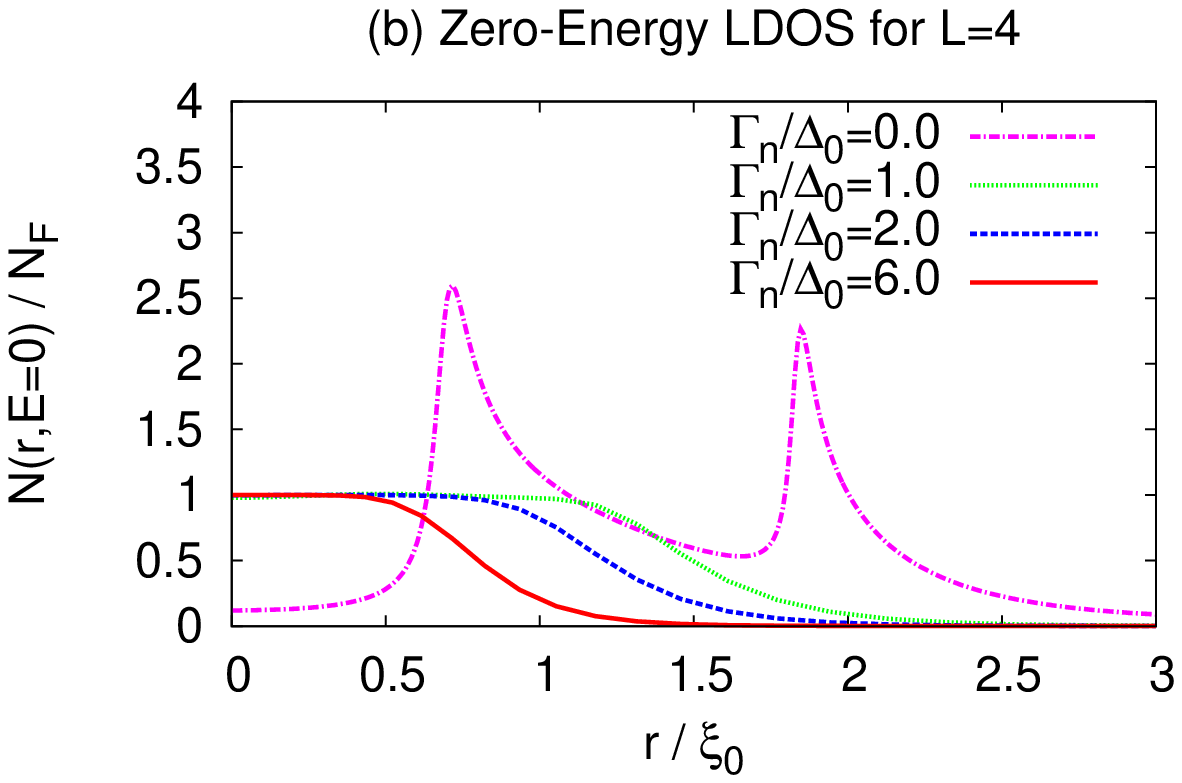}
\end{figure}


\end{document}